# BOLD PSF: Impact of k-space trajectories and timing on the depiction of $T_2$* contrast

Maria Engel[1], Lars Kasper[1,2], Klaas P Pruessmann[1]

[1]Institute for Biomedical Engineering, ETH Zurich and University of Zurich, Zurich, Switzerland

[2]Translational Neuromodeling Unit, IBT, University of Zurich and ETH Zurich, Zurich, Switzerland

**Corresponding author**: Maria Engel, Institute for Biomedical Engineering, ETH Zurich and University of Zurich, ETZ F 87, Gloriastrasse 35, 8092 Zurich, Switzerland. E-mail: engel@biomed.ee.ethz.ch

**Institution**: Institute for Biomedical Engineering, ETH Zurich and University of Zurich, Zurich, Switzerland

# 1 Introduction

Blood-oxygenation-level-dependent (BOLD) fMRI rests on the detection of slight differences in the behaviour of MR signal decay over time. Under the assumption of a mono-exponential decay, BOLD contrast first increases steeply, reaches its maximum at the time constant $T_2^*$, and drops gently thereafter [1]. For that reason, the timing of conventional BOLD fMRI experiments is chosen such that the k-space centre and thereby TE coincides with the peak of the BOLD contrast curve as it manifests for grey matter (GM) at the respective field strength [2].

Spiral readout schemes exhibit more efficient k-space coverage than Echo-planar imaging (EPI) with regard to the exploitation of gradient system constraints. However, the exclusive focus on TE for the choice of timing holds back spiral capabilities in BOLD fMRI. It induces an overly long dead time for outward spiraling trajectories and a resolution-limited inflexibility of timing for inward going spirals.

The argument for the conventional timing relies on the assumption of BOLD contrast being encoded in the centre of k-space. This is a good approximation when the acquisition time is shorter than the timescale on which BOLD contrast changes significantly. Furthermore, it is a reasonable perspective if the k-space centre is sampled in the centre of the acquisition window, as it is the case for EPI. Yet, it does not fully capture the situation when extended readouts are deployed, which sample the k-space centre at any one time within the acquisition window. Like the image itself, the contrast is determined by contributions from all of k-space.

In the present work, the impact of this trajectory-dependent mixture on different characteristics of the BOLD contrast is investigated. This is done by introducing the BOLD PSF as the Fourier transform of BOLD contrast distribution in k-space. It characterizes the imaging process and is different from previous notions of PSF in BOLD imaging, describing the translation of neuronal activation to oxygenation change of haemoglobin [3–7] and also from the PSF of general gradient-recalled-echo (GRE) imaging [8,9].

The BOLD PSF is computed for different trajectories, namely EPI, spiral-out and spiral-in, for different values of $T_2^*$ and with varying TE. Specific metrics are defined to compare sensitivity, resolution and specificity of the respective imaging processes as conveyed by the PSFs.

# 2 Methods

## 2.1 BOLD PSF: Theory

A typical measure of depiction quality in imaging experiments is the point spread function (PSF), describing the response of the imaging system to an input point source[10]. For linear shift-invariant systems, the image of an extended object is the convolution of the true object with the PSF.

In MR Fourier encoding, the PSF is the inverse Fourier transform of the k-space filter $\mathcal{H}(\boldsymbol{k})$ [11]

$$PSF(\boldsymbol{x}) = \mathcal{F}^{-1}(\mathcal{H}(\boldsymbol{k})) \qquad [1]$$

with $\boldsymbol{k} = [k_x\ k_y\ k_z]^T$. The k-space filter $\mathcal{H}(\boldsymbol{k})$ comprises the finite sampling support, as well as any signal weighting that is superimposed.

$$\mathcal{H}(\boldsymbol{k}) = \mathcal{H}_{support}(\boldsymbol{k}) \cdot \mathcal{H}_{weight}(\boldsymbol{k}) \qquad [2]$$

Here, we omit the fact that k-space is sampled at discrete points, assuming that the effect is compensated by an appropriate density correction prior to the image reconstruction.

In GRE experiments, a mono-exponential $T_2^*$ decay is usually assumed [12] such that the signal weighting reads

$$\mathcal{H}_{weight}(\boldsymbol{k}) = \mathcal{H}_{T_2^*}(\boldsymbol{k}) = e^{-\frac{t(\boldsymbol{k})}{T_2^*}} \qquad [3]$$

and provokes a broadening of the main lobe of the PSF along the phase-encoding (PE) direction for EPI and in the radial direction for spirals. Examples are displayed in Figure 1.

In a BOLD experiment, the interesting output is not the images themselves but rather their differences, originating in temporal $T_2^*$ changes. Therefore, BOLD imaging is characterized by a differential PSF.

$$\begin{aligned} PSF_{BOLD}(\boldsymbol{x}, T_2^*, \Delta T_2^*) \\ &= PSF_{GRE}(\boldsymbol{x}, T_2^* + \Delta T_2^*) - PSF_{GRE}(\boldsymbol{x}, T_2^*) \\ &\approx \frac{\partial PSF_{GRE}(\boldsymbol{x}, T_2^*)}{\partial T_2^*} \Delta T_2^* \end{aligned} \qquad [4]$$

using the Taylor approximation for small $\Delta T_2^*$ in the last step. Since the Fourier transform is a linear operation, we may interchange it with the derivative of the GRE imaging PSF with respect to $T_2^*$ and as the sampling filter does not depend on $T_2^*$, only the weighting function must be adapted:

$$\mathcal{H}_{weight}^{BOLD}(\boldsymbol{k}) = \frac{\partial \mathcal{H}_{T_2^*}(\boldsymbol{k})}{\partial T_2^*} = \frac{t(\boldsymbol{k})}{{T_2^*}^2} e^{-\frac{t(k)}{T_2^*}} \quad [5]$$

It is maximized at $t(\boldsymbol{k}) = T_2^*$, which is the reason that this time is commonly chosen for TE. However, this disregards the BOLD weighting elsewhere in k-space, which is different especially for extended readout trains (Figure 1, centre).

The full PSF of the BOLD imaging part then reads

$$PSF_{BOLD}(\boldsymbol{x}, T_2^*, \Delta T_2^*) = \mathcal{F}^{-1}\left(\mathcal{H}_{support}(\boldsymbol{k}) \cdot \frac{t(\boldsymbol{k})}{{T_2^*}^2} e^{-\frac{t(k)}{T_2^*}}\right) \Delta T_2^* \quad [6]$$

Because $T_2^*$ changes increase with field strength, but are also paradigm-dependent, normalization of Equation $[6]$ by relative $T_2^*$ change ($\Delta T_2^*/T_2^*$) is here chosen as measure for comparison:

$$PSF_{BOLD}(\boldsymbol{x}, T_2^*, \Delta T_2^*) = \mathcal{F}^{-1}\left(\mathcal{H}_{support}(\boldsymbol{k}) \cdot \frac{t(\boldsymbol{k})}{T_2^*} e^{-\frac{t(k)}{T_2^*}}\right) \quad [7]$$

Like the GRE PSF, the BOLD PSF is not shift-invariant due to its dependence on $T_2^*$, which is spatially variable. However, it may be assumed that in a BOLD experiment at a given field strength, the interesting signal changes stem from regions with similar $T_2^*$.

## 2.2 Conceptual Consequences

The described BOLD PSF concept entails two consequences that can already be observed qualitatively in Figure 1. Firstly, the BOLD PSF differs from the GRE PSF in different manner depending on the k-space trajectory. In case of spiral out, in particular, it has a narrower main lobe and thus suggests a *higher* resolution for BOLD contrast than for $T_2$* contrast in GRE imaging. Secondly, different trajectories can exhibit a similar BOLD signal despite a very different nominal TE, as it is the case for spiral-out and spiral-in here. Examples of varied timing and resulting BOLD PSFs are displayed in Figure 2.

## 2.3 Comparative approach

We computed the BOLD PSF according to Equation 7 for 2D EPI and spiral readout schemes. The input $t(\boldsymbol{k})$ depends on the start time of the readout and the exact trajectory design. The latter was here computed for realistic examples of different resolutions. In each case, the covered k-space area and the readout duration were matched between EPI and spiral such that the spiral had a slightly lower undersampling factor. Note that this choice is tailored to provide a fair PSF comparison but is not suited for SNR comparisons since spirals usually qualify for higher undersampling factors by virtue of distributing noise amplification from parallel imaging more evenly in k-space due to a more balanced distribution of k-space velocities. All trajectories were computed so as to minimize their duration within gradient-strength and slew-rate constraints of common gradient systems ($G_{max}$ = 31 mT/m, $S_{max}$ = 200 T/m/s) [13]. $t(\boldsymbol{k})$ was then gridded onto a rectangular grid using a sufficiently wide kernel to make up for the undersampling. In this way, potential artifacts and noise amplification arising from undersampling or oversampling are disregarded. The support filter $\mathcal{H}_{support}(\boldsymbol{k})$ was modeled as a rectangular support for EPIs and a circular support for spirals with the same area and in accordance with the respective trajectories.

## 2.4 Quality metrics

Since all k-space filters under investigation are real-valued, any phase variation in the resulting PSFs is symmetric about the centre of the main lobe and will cancel out. Therefore, further analysis of the PSFs was performed using the real part only [6,14]. It was assessed through three different measures, all of which take into account the two dimensions of the PSF:

1. *Resolution*: The FWHM of the main lobe, computed as the diameter of a disk with area equal to the area that appears at half height of the main lobe. This approach was taken to ensure comparability between spirals and EPI despite the latter exhibiting anisotropic PSFs.
2. *Sensitivity*: The integral over the main lobe inside an isoline of 20 % of its peak value. The choice of the boundary line distinguishing between signal that does and signal that does not belong to the main lobe is heuristic but limited by the fact that not all of the investigated PSFs have a zero crossing in all directions between the main lobe and the first side lobe.
3. *Specificity*: The relative side lobe suppression as the relation between the integral over the main lobe and the $L^2$-Norm of the side lobes. In analogy to the definition of SNR[15], this definition is driven by the consideration of the main-lobe as correct signal contribution and of the side lobes as noise-like nuisance.

## 2.5 Imaging scenarios

This procedure was repeated for TEs up to 100 ms with a step size of 4 ms. Piecewise polynomials and second order exponentials were fitted to the resulting TE-dependent resolutions, sensitivities and specificities (measures 1-3) to analyse the curves further, i.e. to compare TE, resolution and specificity for maximized and levelled sensitivities.

We aimed to cover the following imaging scenarios:

A. **HCP 3T:** Resting state fMRI according to the protocol of the Human brain project (HCP) with 2 mm resolution, Multiband factor 8, no in-plane acceleration, and $T_2^*$ = 66 ms, reflecting GM at 3T[16]. To imitate the original protocol, the gradient system specifications were reduced to $G_{max}$ = 30 mT/m, $S_{max}$ = 180 T/m/s.
B. **High-Res 3T:** High-resolution fMRI at 3T with 1mm resolution, R = 4 in-plane acceleration only.
C. **Laminar 7T:** Laminar fMRI at 0.8mm resolution, R = 4 in-plane acceleration only, and $T_2^*$ = 33.2 ms, reflecting GM at 7T.
D. **Resolution Sweep 3T and 7T:** A range of 23 different resolutions between 0.8 mm and 3 mm, R = 4 in-plane acceleration only, $T_2^*$ = 66 ms, reflecting GM at 3T and $T_2^*$ = 33.2 ms, reflecting GM at 7T.

Note that despite the Multiband scenario (A) involving a third encoding dimension, there is no filtering effect in that dimension for the given trajectories and it suffices to restrict the analysis to two dimensions.

# 3 Results

Figure 3 shows the three BOLD PSF measures for spiral-out, spiral-in, and EPI in scenarios A, B and C. Spirals have a 2-3% higher BOLD sensitivity peak than EPI. This holds true for all three scenarios and both, spiral-in and spiral-out, except for the spiral-in in the Laminar fMRI case, where the shortest possible TE is 64 ms. The optimum BOLD sensitivity is reached at earlier TEs for spiral-out than for EPI and at later TEs for spiral-in trajectories. Resolution and specificity always show opposing behaviour, so that higher resolution goes along with reduced specificity – due to the higher contamination of other locations via PSF side lobes – and vice versa. For spiral-out, the resolution decreases with increasing TE. This is expected because when TE increases, the maximum BOLD contrast at $T_2^*$ shifts to lower k-space frequencies. In contrast, resolution of EPI and spiral-in increases with TE, with EPI curves being, in general, flatter.

Figure 4 displays the results of the study on resolution dependence of the BOLD PSF ("Resolution Sweep 3T"). A and B illustrate how optimizing TE and sequence timing with respect to the introduced sensitivity of the BOLD PSF depends on the trajectory and changes with the chosen nominal resolution. A confirms that for EPI, the optimum TE is close to $T_2^*$ (indicated by the dashed black line), as in traditional theories of BOLD sensitivity [17–19] evaluating k-space centre only. This holds true across all resolutions with a slight but increasing deviation for higher resolutions – reflecting the asymmetry of the filter $\mathcal{H}_{weight}^{BOLD}$, which bears more impact on long readout trains. On the contrary, substantial deviations from $TE = T_2^*$ being optimal occur for both spirals and in opposite directions. Again, these deviations are greater for higher resolutions due to longer readout trains. Since both, spiral-out and spiral-in, reach higher maximum sensitivities than EPI, levelling the sensitivity with the maximum BOLD sensitivity of the respective EPI allows for shorter echo times (light blue and orange curves). For spiral-out, this leads, depending on the resolution, to up to 61% decrease in TE compared to the $TE = T_2^*$ approach.

Given these timing choices, Figure 4 B illustrates the resulting minimum repetition times. Due to shortening readout durations, the minimum TR decreases for all trajectories with coarser resolutions. Timing the readout of a Spiral-out trajectory so that it maximizes the BOLD sensitivity (blue curve) requires across all resolutions a 1%-12% longer slice-TR (3%-19% for 7T, see Figure 5) than EPI (green curve). However, levelling the sensitivity provides a 2% - 17% shorter slice-TR (light blue vs green curve). For 7T, this figure reads 14% maximum at medium and low resolutions and reverses for higher than 1 mm resolution, suggesting up to 7% longer slice-TRs. For spiral-in on the other hand, choosing the maximum BOLD sensitivity comes along with a 2% - 12% reduced slice-TR (red curve) (2% - 16% at 7T). Levelling the BOLD sensitivity of spiral-in with EPI's maximum BOLD sensitivity provides even more, namely a 19% - 26% decrease in slice-TR (orange curve) (5% - 27% at 7T). Taken together, independent from resolution, the highest BOLD sensitivity at shortest minimum TR can be accomplished with spiral readout schemes.

Figure 4 C and D illustrate the dependence of BOLD resolution and specificity on nominal resolution for the same trajectories and timing choices as they are underlying A and B. For spiral-in, levelling BOLD sensitivity to the EPI maximum and thereby reducing dead time in the sequence, sacrifices BOLD resolution. The opposite is true for spiral-out: levelling the sensitivity improves the BOLD resolution. Note that the reason for this effect largely lies in the fact that the sensitivity-levelling effect can be reached either by pulling the spiral trajectories forward or by delaying them and that the former was chosen here since they bear the additional benefit of minimizing slice-TR.

Apart from very low resolutions, spiral-out shows a slightly higher BOLD resolution than spiral-in when operating at maximum BOLD sensitivity. Across all resolutions, EPI shows a higher BOLD resolution but worse specificity than both spirals at their maximum and EPI-levelled sensitivity.

Some of these observations are opposing the results derived from traditional GRE PSF evaluations: To illustrate that, additionally, resolution and specificity of the respective GRE PSFs are depicted in the same plots for the scenarios maximizing BOLD sensitivity. Spiral-in and EPI exhibit a worse BOLD resolution than their GRE resolution, while spiral-out trajectories exhibit the opposite effect

# 4 Discussion

The proposed BOLD PSF concept is a tool to assess the qualification of arbitrary readout trains for, and their impact on, fMRI BOLD contrast. It suggests that the choice of TE in such experiments is subject to complex optimization, which needs to involve not only the sensitivity of the BOLD signal as objective function but also its resolution and specificity. As such, it goes beyond (i) the notion of a PSF of the GRE imaging underlying the BOLD experiment, as well as (ii) a purely TE-based BOLD sensitivity, and it alters the results of such considerations.

The presented results suggest that in general, $T_2$* is not necessarily the optimum TE in a BOLD experiment. Instead, the chosen timing should depend on the sampling strategy, the length of the readout train, the field strength in question, and, ultimately, the target application. The basis of this lies in viewing the BOLD contrast curve less as a guide to optimize the acquisition time of the k-space centre, but rather as a BOLD image filter corresponding to a BOLD PSF. The discrepancy between the traditional perspective and the new concept becomes more pronounced for higher field strengths and for longer readout trains.

The most pivotal result of this approach is the fact that for spiral-out, the echo time should always be chosen smaller than $T_2$*. This is true if maximum BOLD sensitivity ought to be reached. Notably, this maximum is always higher for spiral-out than for EPI. Further shortening of TE may be indicated for application-specific balancing between BOLD sensitivity, resolution and specificity.

For spiral-in, the optimization of timing gets slightly more intricate. Unlike spiral-out, it delivers the maximum BOLD sensitivity only for TE > $T_2$*, which alleviates the requirements imposed on the readout duration to fit before TE. Spiral-in seems on par with spiral-out over a large range of the studied parameter space, regarding the maximum BOLD sensitivity, and thus, considering its shorter minimum TR, even more attractive. This is except for high resolutions at high field strengths, where the combination of long readout trains and short $T_2$* values implies that the optimum TE cannot be reached anymore. In this very same realm, spiral-out – although still delivering the highest possible sensitivity – forfeits it attractiveness as even the version with levelled sensitivity requires a minimum slice-TR that is longer than the one achieved by EPI. Hence, EPI might seem like the best choice regarding the minimum slice-TR at fix sensitivity for this combination of parameters. However, a final conclusion cannot be drawn without considering SNR, which was neglected in this work.

If the readout length allows, the spiral-in could be shifted forward *and* backward with respect to the sensitivity-maximizing TE without falling behind EPI in terms of minimum slice-TR and BOLD sensitivity (parameter space between the orange and green curve in Figure 4B). Although Figures 4 and 5 only display the forward shifting case to minimize slice-TR, it is straight-forward to also calculate the backward one. Both will deliver different trade-offs between BOLD resolution and specificity due to the asymmetry of the filter $\mathcal{H}_{weight}^{BOLD}$ and thereby provide additional flexibility.

As mentioned earlier (2.2), the shown framework demonstrates that the BOLD PSF can be in different ways more or less favourable than the respective GRE PSF, depending on readout type and timing. Wherever the BOLD PSF exhibits a smaller main lobe than the GRE PSF, as it is the case for spiral-out, this can, with the equivalency of PSF and SRF, be interpreted as signal from a narrower region contributing constructively to the signal in a given voxel. Both, spiral-in and EPI show the opposite effect.

The shown curves hold true for a specific gradient system as can be found in a typical clinical context nowadays. Recent advances in the realm of high-performance gradient systems[20–24] raise the question how such systems would impact the results. In general, under the assumption that such devices allow for faster sampling of the same k-space area at fix sampling density, i.e. a higher average speed in k-space, all of the shown curves in Figures 4 and 5 are expected to shift to the left (in the direction of smaller voxels). This means that at a given nominal resolution, the differences in optimum TE between different trajectories become smaller. Furthermore, in the parameter space of nominal resolution and field strength, the domain where spiral-in does not reach its BOLD sensitivity maximum anymore is pushed towards finer resolution and higher fields. Alternatively, stronger gradient systems may be used to increase the SNR by reducing the undersampling. Which of the options – higher resolution, faster scanning, or less noise amplification – is preferable in a given imaging situation is again subject to more comprehensive considerations of SNR, including coil profiles and tissue characteristics.

In general, it should be noted that retrospective balancing of resolution and specificity with sensitivity remaining unchanged is always feasible with suitable high-pass or low-pass filters.

Computations were here performed under two simplifying assumptions. On the one hand, a mono-exponential signal decay does not describe the relaxation process in brain tissue entirely accurately [25,26]. Yet, the framework may be applied to any other more comprehensive relaxation model (e.g. [27]). On the other hand, the simple Fourier picture, from which we started, does not suffice to describe the entire imaging process. In reality, also receive coil footprints will add to the shape of the PSF [28] as well as static and dynamic spatial field inhomogeneities, implying, in particular, that the assumed shift-invariance does not hold[28,29]. These aspects were not included in the current theoretical work in order to focus on the impact of k-space sampling strategies alone.

In actual experiments, SNR will be an important criterion for the choice of timing and trajectory, apart from the mentioned three measures, BOLD sensitivity, specificity and resolution. Since spiral sampling provides a more even distribution of g-factor noise amplification in the image than EPI [30], SNR considerations are expected to further strengthen its case. This expected strengthening is firstly due to a more even distribution of undersampling in all directions and thereby better exploitation of the encoding capabilities of multi-dimensional coil arrays. Secondly, it is due to the comparison strategy based on duration-matching that was pursued here, and which resulted in lower undersampling factors for spirals than for EPI since spirals use the capabilities of common gradient systems more efficiently. The added efficiency of spirals could likewise be invested into higher resolution.

Taken together, spiral sampling seems with respect to the BOLD PSF at least coequal to EPI, however at the price of image reconstruction complexity and the inability to realize anisotropic sampling such as Partial Fourier Imaging [31].

In this work, the impact of k-space sampling on $T_2$* contrast was studied with the focus on BOLD fMRI. However, the theory should hold true also for other applications such as Dynamic Susceptibility Contrast (DSC) perfusion imaging [32] or spin-echo sequences with an accordingly modified signal filter. In both cases, the decisive factor of signal decay occurring on a similar time scale as the readout duration is given.

With*in* the BOLD imaging context, it is important to note that the validity of the concept is orthogonal to other PSF concepts, e.g. describing the transfer of neuronal activation to BOLD response. In particular, it is separate from and therefore compatible with considerations regarding the specificity of the BOLD response being subject to caveats investigated by CBF and CBV fMRI [33].

Whereas the present work focuses on the theoretical foundations, future work should investigate the proposed concept in in-vivo experiments to corroborate the current findings.

# 5 Acknowledgments

The authors would like to thank Jakob Heinzle for fruitful discussions.

# 7 Figures

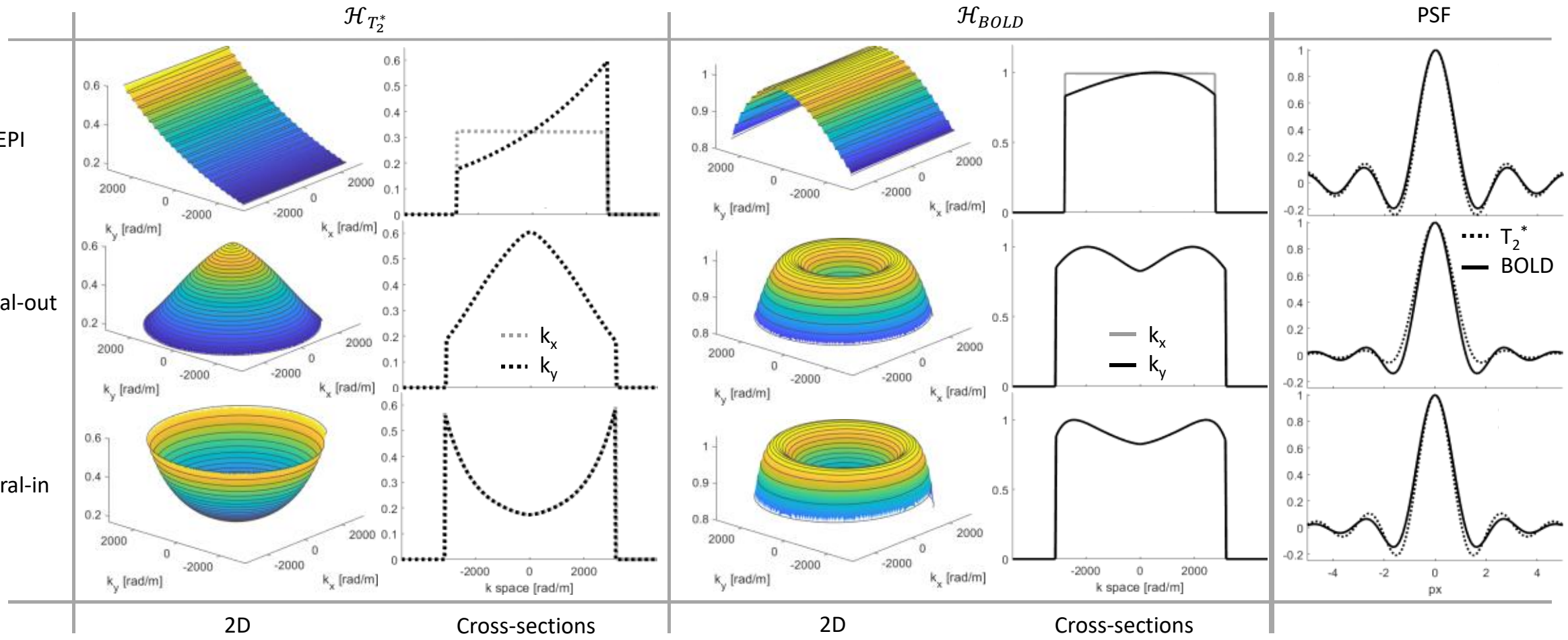

**Figure 1**: GRE and BOLD filter in k-space for different trajectories and the corresponding PSFs. In this timing example, the readout is symmetrically arranged about $T_2$*. $\mathcal{H}_{T_2^*}$ is normalized to the initial signal amplitude, $\mathcal{H}_{BOLD}$ and the PSFs are normalized to their respective maximum.

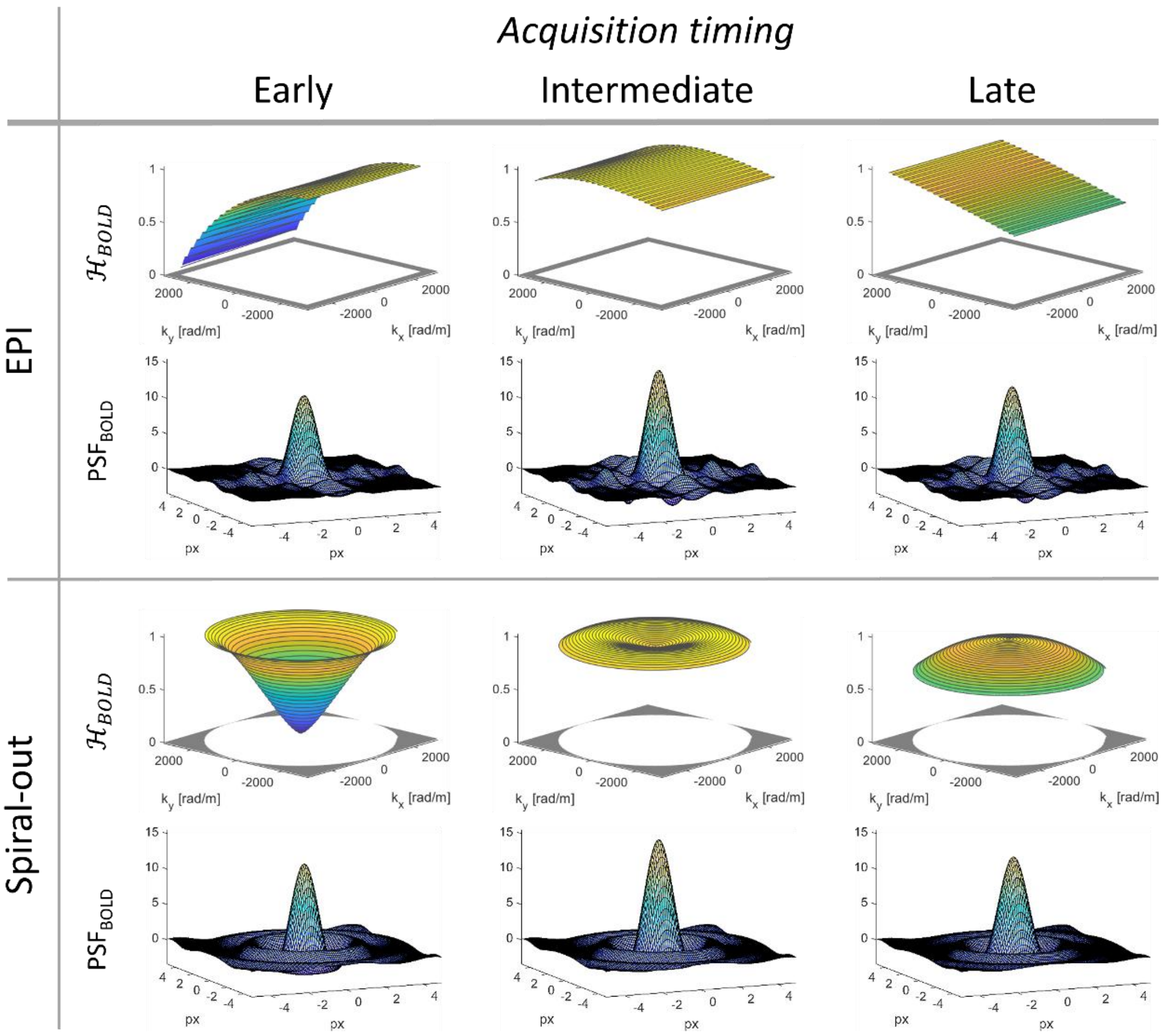

**Figure 2**: Impact of the timing on BOLD imaging filter and PSFs. Left column: Readout starts long before $T_2$*. Central column: Readout is arranged symmetrically about $T_2$*. Right column: Readout starts later than $T_2$*. Since TE is defined as the time of the acquisition of k-space centre, spiral-out exhibits in each column a shorter TE than the respective EPI.

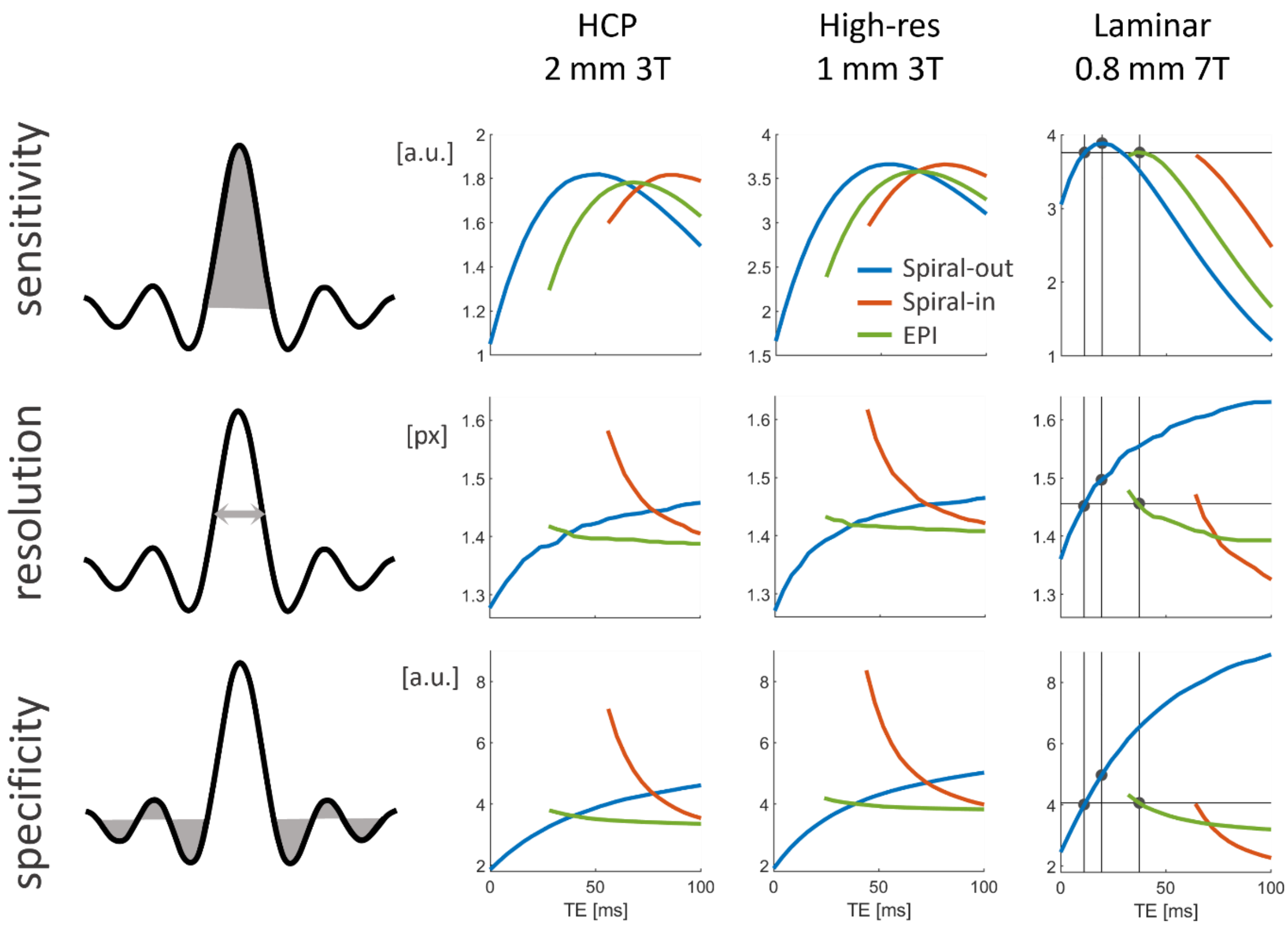

**Figure 3:** Sensitivity, resolution and specificity of the BOLD PSF for three different typical use cases: Firstly, an imaging scenario as it was defined for the Human Connectome Project, secondly a higher resolution scenario at 3T, and lastly a sub-mm resolution scenario at 7T as typically devised for experiments seeking to resolve cortical layers. For the latter one, vertical gray lines exemplarily indicate, at what TE maximum BOLD sensitivities are reached for EPI and Spiral-out and at which shorter TE Spiral-out trajectories deliver the same BOLD sensitivity as the maximum for EPI. Gray spheres mark this sensitivity, as well as the resolution and specificity reached at the respective echo times. Vertical gray lines at the respective values for EPI facilitate comparison of the figures.

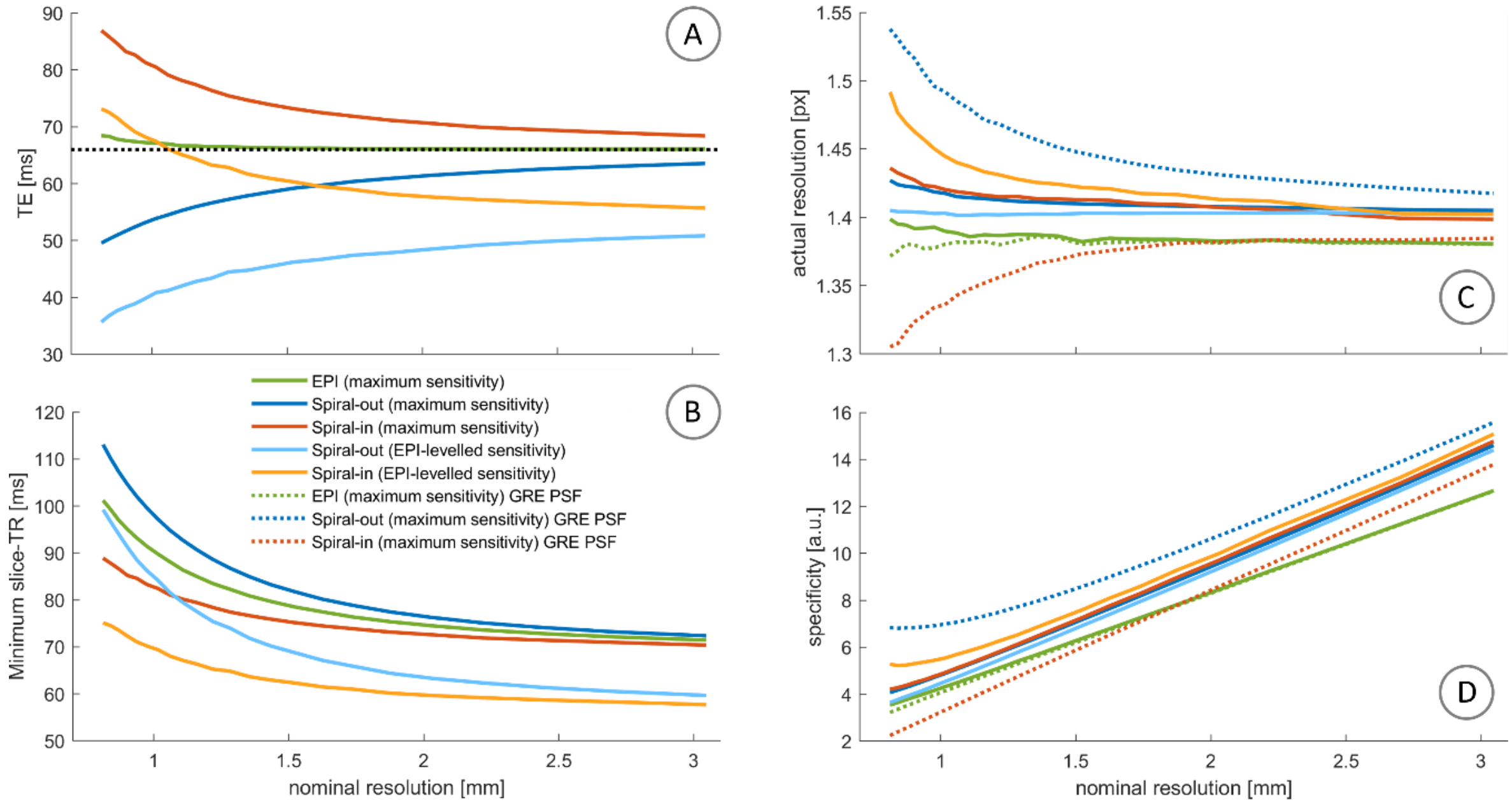

**Figure 4**: A: For all three trajectory types (EPI, Spiral-out and Spiral-in), the TE which maximizes the BOLD sensitivity is plotted against the nominal resolution. Additionally, for both spirals, the shortest TE is plotted for which the BOLD sensitivity is on par with the respective EPI BOLD sensitivity. The dashed black line indicates $T_2^*$. B: Resulting minimum slice-TR (TE chosen as in A). It is computed as the extent of the readout after the centre of the excitation pulse (zero point of the time axis) plus 2 ms for the first half of the excitation pulse. Right: Solid lines indicate BOLD resolution (C) and BOLD specificity (D). Dashed lines indicate the respective resolution and specificity of the GRE PSF.

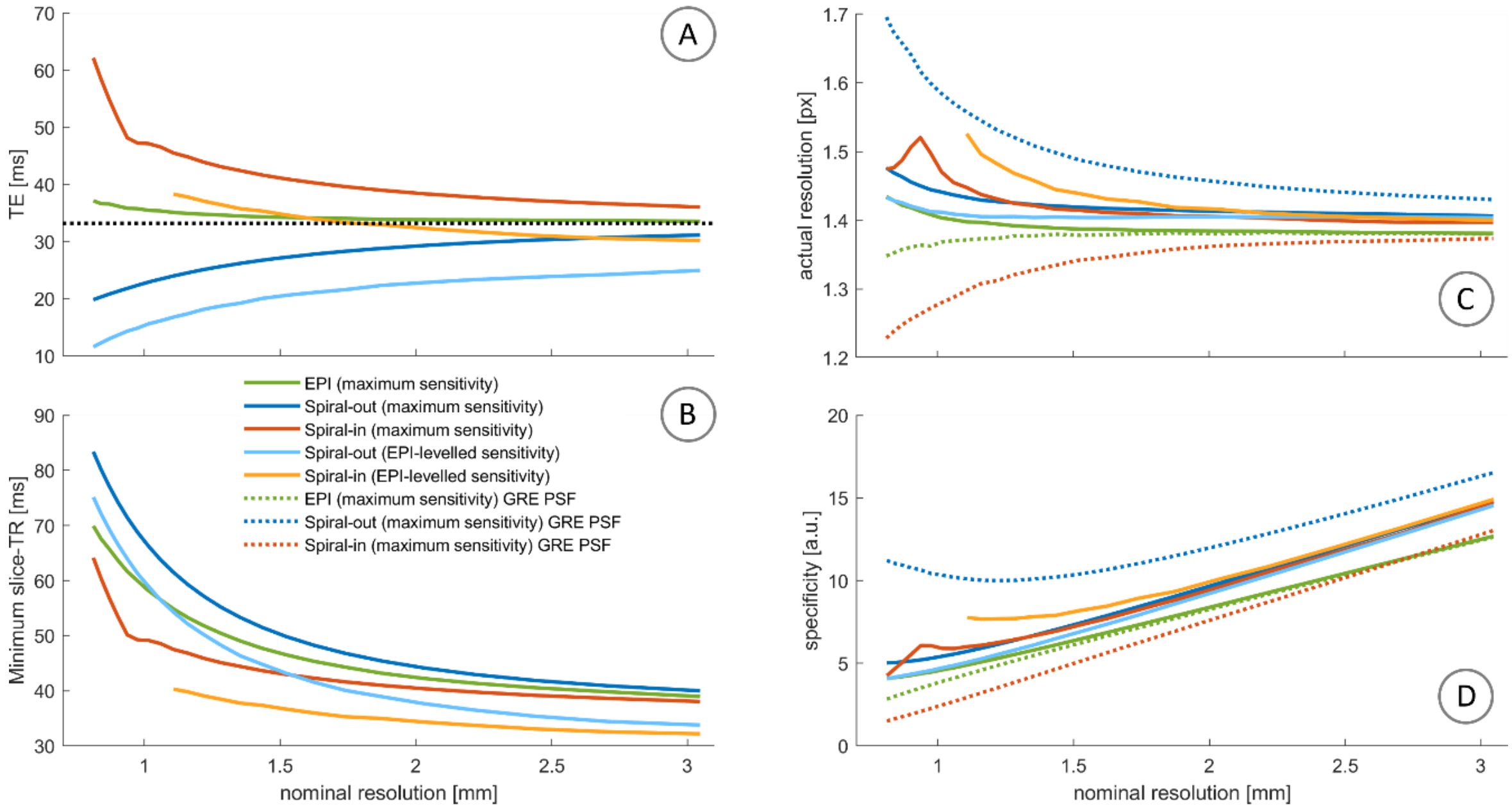

**Figure 5**: Results from Resolution Sweep (see Figure 4) for 7T. For high spatial resolutions, the shortest possible TE that can be reached with Spiral-in is longer than what would be needed to level the BOLD sensitivity the one of EPI by shifting the readout to the front. Therefore, the orange curve is not plotted for nominal resolutions finer than 1.1mm. For nominal resolutions finer than 1mm, the BOLD sensitivity does not reach a maximum anymore.